\documentclass[12pt,preprint]{aastex}

\newcommand{\AaA}{A\&A}

\newcommand{\ApJ}{ApJ}
\newcommand{\MNRAS}{MNRAS}

\newcommand{\um} {$\mu$m}

\newcommand{\simless}{\mathbin{\lower 3pt\hbox
      {$\rlap{\raise 5pt\hbox{$\char'074$}}\mathchar"7218$}}} 
\newcommand{\simgreat}{\mathbin{\lower 3pt\hbox
     {$\rlap{\raise 5pt\hbox{$\char'076$}}\mathchar"7218$}}} 

\shorttitle{Diamonds in HD~97048}
\shortauthors{Habart et al.}

\begin{document}


\title{Diamonds in HD~97048: a closer look}


\author{E. Habart, L. Testi, A. Natta, M. Carbillet\altaffilmark{1}}

\affil{INAF--Osservatorio Astrofisico di Arcetri, Largo E.\,Fermi 5,
           50125 Firenze, Italy}
\email{habart@arcetri.astro.it, lt@arcetri.astro.it, natta@arcetri.astro.it, marcel@arcetri.astro.it}


\altaffiltext{1}{On move to Laboratoire Universitaire d'Astrophysique de Nice, UMR 6525 du CNRS, Parc Valrose, 06108 Nice Cedex 2, France}


\begin{abstract}
We present adaptive optics high angular resolution ($\sim$0$\farcs$1) spectroscopic observations in the 3 $\mu$m
region
of the Herbig Ae/Be star HD~97048.
For the first time, we spatially resolve the emission
in the diamond features at 3.43 and 3.53 $\mu$m and in the adjacent continuum.
Using both the intensity profiles along the slit and reconstructed two-dimensional images of the object, we derive full-width at half-maximum sizes consistent with the predictions for a circumstellar disk seen pole-on.
The diamond emission originates in the inner region
($R \lesssim 15$ AU) of the disk.
\end{abstract}





\section{Introduction}

HD 97048 is a Herbig Ae/Be star, member of the Chamaeleon~I association, situated at 180~pc
\citep{vandenancker97,whittet97}.
\citet{the86} derived from a photometric and spectroscopic study of
HD~97048 that its infrared (IR) excess  originates in a flattened shell
structure or disk of dust particles seen pole-on.
The unresolved SEST image of HD~97048 taken by \citet{henning98} together with the measured mm-flux also points towards the presence of a flattened disk-like structure \citep{henning93,henning98}.
The object is not extended in the near-IR on scales of $\lesssim 1\arcsec$ \citep{henning98}, but it is in the mid-IR on scales of 5-10\arcsec\ \citep{prusti94,siebenmorgen2000}.
The mid-IR emission, dominated by aromatic infrared emission bands \citep{siebenmorgen2000}, is most probably due to an extended envelope  of transiently heated
very small grains and Polycyclic Aromatic Hydrocarbons (PAHs).
However, \citet{boekel03} have recently spatially resolved the  10 $\mu$m emission of  the central source,
showing that most of the emission in the
8.6, 11.3 and 12.7 $\mu$m features comes in fact
from a region of 1.2-1.6$\arcsec$ or 200-300 AU, very likely a disk.

The spectra of HD~97048 show, next to the infrared emission bands at 3.3 \um\ attributed to PAHs, the rare
emission features at 3.43 and 3.53 \um\ \citep[e.g.,][]{blades80,guillois99,vankerckhoven2002}.
\citet{guillois99} proposed to attribute
these bands to diamond surface C-H stretching features; an identification which, because of the good match
between laboratory and observed spectra, is very persuasive. Nanodiamonds are  the most abundant
presolar grains in primitive meteorites, and they could in fact constitute an important component of the
dust in circumstellar environments and in the interstellar medium \citep[e.g.,][]{jones2004}.
However, the diamond particles in HD~97048 or Elias~1 are probably much larger (size of $\sim$10--50\,nm or
larger) than the ``true'' nanodiamonds (mean sizes of the order of 2--3\,nm) observed in primitive
meteorites \citep{jones2004}.

One way to better understand the properties and evolution of diamond particles is to study the spatial
distribution of their emission. Presently, the only observational constraints one has
come from speckle observations by \citet{roche86}
who found that the emission in the 3.53 \um\ band in HD~97048 is unresolved and arises from
a region $<$0$\farcs$1 in diameter.
Here, we present adaptive optics high angular resolution ($\sim$0$\farcs$1) spectroscopic observations of
HD~97048.
These observations allow us  to spatially
resolve for the first time the 3 \um\ features attributed to diamonds, as well as the adjacent continuum.
In Sect.\,2, we present the observations. The results are presented in Sect. 3
and discussed in Sect. 4.

\section{Observations and data reduction}

Long slits spectra in the L-band (3.2--3.76 \um) of HD~97048 were taken on 10 January 2003 with the
adaptive optics (AO) system NAOS-CONICA (NACO) at the VLT, using the visible 14$\times$14 Shack-Hartmann wave-front
sensor working at nearly 500\,Hz. The 28$\arcsec$ long slit had a width of 0$\farcs$086 which roughly corresponds
to the diffraction limit while the pixel scale was $\sim$0$\farcs$0547
and the spectral resolution $\sim$700.
We took 7 slit positions: one centered on the star and the others at 0$\farcs$043, 0$\farcs$086,
0$\farcs$129 and -0$\farcs$043, -0$\farcs$086, -0$\farcs$129 off axis.
For each slit position, in order to correct from the atmospheric and instrumental background, we employed
standard chop/nod technique with a throw of $\sim$9\arcsec\ in the north-south direction.
 The integration time per chop- or nod-positions was 1~min.

The log of the observational conditions reports a Fried parameter $r_0$
close to 10\,cm (at 500\,nm), an average outer-scale ${\cal L}_0$ of $\sim$15\,m, and a resulting
average coherent energy (after AO correction and at 2.2 \um) of up to 40\% on axis. The latter roughly gives the
actual Strehl ratio characterizing the quality of our data. According to 
simultaneous measurements, the  seeing (at 500\,nm) was $\sim$1$\farcs$1 consistent
with
the NACO values. Our observations clearly benefit from good seeing conditions and AO
correction, and the achieved angular resolution was close to the diffraction limit.

Data reduction was performed using firstly a standard procedure for IR spectroscopic observations
(pre-processing phase), and then a classical Richardson-Lucy (LR) deconvolution method.
During the pre-processing phase, to remove the telluric features we used the observations of a spectroscopic
standard star (HIP 53074) taken on 20 February 2003, with the same airmass.
For the deconvolution procedure,
and to have a suitable estimate of the point-spread function (PSF), we used instead
a second reference star (HIP 20440), observed
on 19 January 2003, which provided a
better match to our object  in terms of atmospheric turbulence conditions ($r_0$, ${\cal L}_0$,
seeing), and AO correction quality (K-band Strehl ratio $\sim$30\%).

\section{Results}

\subsection{Spectra}

\begin{figure}
\includegraphics[angle=0,scale=.50]{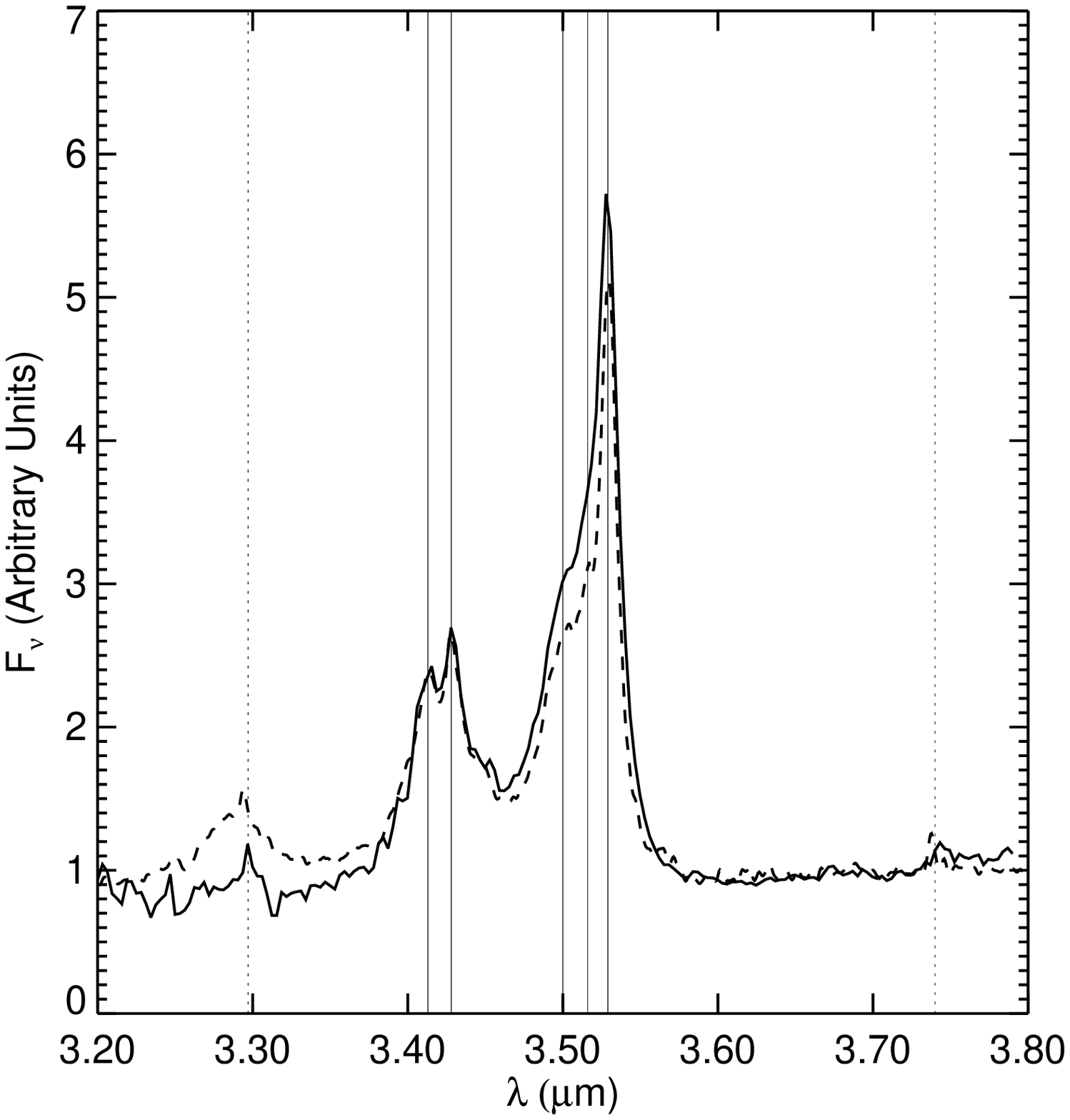}
\caption {Spectrum obtained by combining all the slits and for distances from the star going from 0 to
          1\arcsec\ (solid lines). The dashed line shows the ISO-SWS spectrum,
obtained with a beam of 20$\times$33\arcsec. We also show as solid vertical lines
the diamond feature positions and as vertical dotted lines the position of
the Pf~$\delta$ line (on top of the broad PAH feature at 3.3\um) and the Pf~$\gamma$
line at 3.74 \um.}
\label{spectra}
\end{figure}

Fig.~\ref{spectra} shows the spectrum obtained by combining all the slits and for distances
from the star going from 0 to 1\arcsec. It clearly shows the two strong features at 3.43
and 3.53 \um, attributed to diamonds. The features are not smooth, but show a number of components,
in particular, the 3.41, 3.43, 3.50, 3.51 and 3.53 \um\ ones. These characteristics support the
identification of the carriers of the two features as diamonds \citep[e.g.,][]{guillois99,vankerckhoven2002,jones2004}.
The spectrum shows also a weaker feature at 3.3 \um, characteristic of PAHs. There may be also some
hydrogen emission lines; the Pf~$\delta$ on top of the broad PAH feature at 3.3 \um\ and the Pf~$\gamma$
line at 3.74 \um.

In Fig.\,\ref{spectra}, we also compare our spectrum with that obtained with the 
Short Wavelength Spectrometer (SWS) on board the Infrared Space Observatory (ISO)
 with a slit of 20$\times$33\arcsec. The ISO-SWS spectrum
contains both the disk and emission from the reflection nebula that surrounds the system \citep{siebenmorgen2000,boekel03}.
Comparing the two spectra, we see that the diamond components are very similar in peak position, width and
relative strength. Moreover, the peak/continuum ratio in the diamond bands is roughly similar.
This suggests that diamond emission is confined to the disk. On the other hand, the ISO spectrum shows
stronger PAH emission at 3.3 $\mu$m than the NACO one, consistent with the hypothesis that
the PAH feature seen in the ISO spectrum comes  from both the disk and the surrounding nebula \citep{boekel03}.

\subsection{Brightness spatial distribution}

\subsubsection{Along the slit}

\begin{figure}[htpb]
\includegraphics[angle=0,scale=.50]{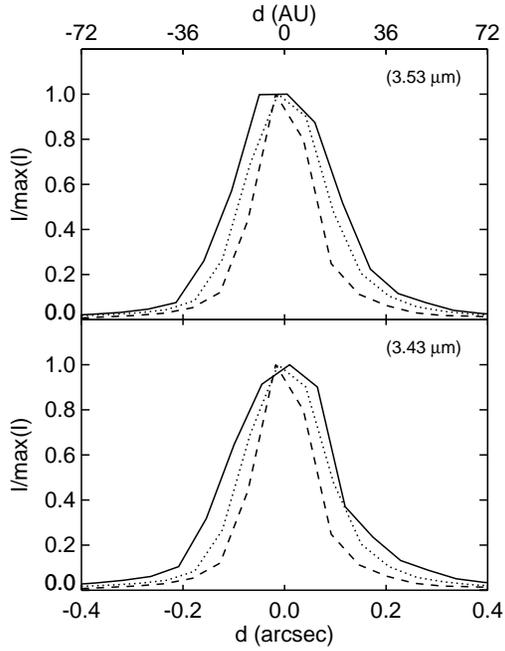}
\caption {Normalized spatial emission profiles of the stronger 3.43 and 3.53 \um\ diamond features (continuum substracted,
          solid lines) as function of the distance from the star. The scale on the lower axis shows
the distance in arcsec, that on the upper axis the corresponding value in AU. Dotted and dashed lines show the
          intensity profile of the continuum and of the PSF, respectively.}
\label{cut}
\end{figure}

\begin{figure}
\includegraphics[angle=0,scale=.90]{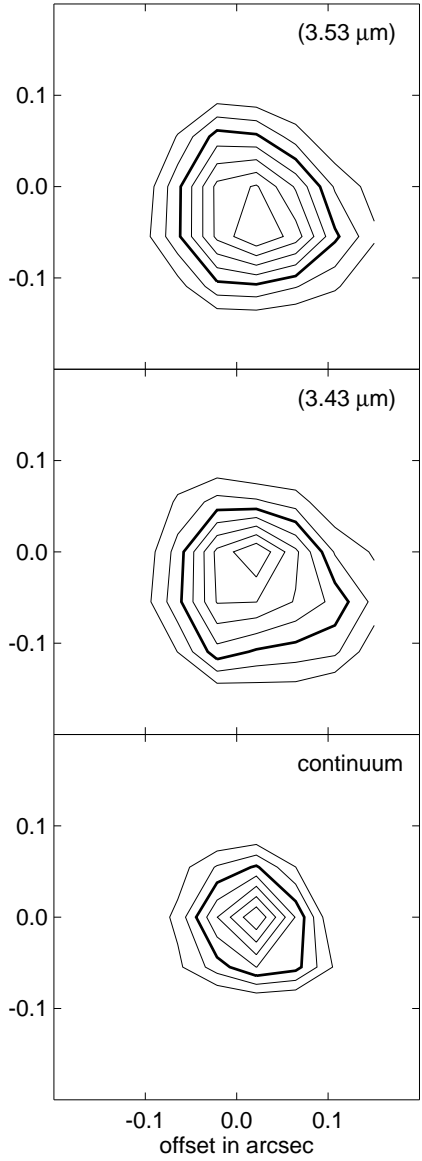}
\caption {Deconvolved normalized emission map of the 3.43 and 3.53 \um\ features (continuum substracted)
          and of the continuum. The contour levels go from 0.3 to 0.9 (step=0.1); the 0.5 level is shown
          in thick line.}
\label{map}
\end{figure}

Fig.~\ref{cut} shows the intensity profile (after continuum subtraction) of the two 3.43 and 3.53 \um\ features
along the slit length.
The intensity profile in the slit centered on the star and at 0$\farcs$043 have been averaged together
because the star was offset by approximately 1/4 of slit (i.e., $\sim$0.02\arcsec, see Fig. \ref{map}).
Fig.~\ref{cut} also shows the intensity profile of the continuum (measured between 3.6 and 3.7 \um)
and of the observed reference star (the assumed PSF).
The spatial extent of the diamond emission in both the 3.43 and 3.53 \um\ features is significantly broader
than that of the continuum, which is slighlty broader, in turn,  than  the PSF.
Note that since the PSF data resulted in a slightly worse Strehl ratio (see Sect. 2) than the
object data, it is unlikely that the apparent extension of the continuum is
due to an additional broadening from the combination of atmospheric conditions and partial AO correction.
The PAH emission appears to be more extended than the diamond features;
it will be discussed in details in a forthcoming paper presenting other NACO observations of the 3.3 \um\ PAH emission band in Herbig Ae/Be stars.

As discussed above, the fact that both the diamond features and the continuum emission
are resolved  is clearly visible already in the pre-processed data.
Nevertheless, in order to better estimate the exact spatial extensions and eliminate, as far as possible, the remaining
atmospheric and post-AO broadening effects on the profiles, we have applied to the data
a classical LR deconvolution method. In practice this was performed using the
Software Package AIRY \citep{Correia2002}. In order to
avoid Gibbs oscillations \citep[also known as ``ringing'' effect -- see][]{Bertero1998} in the rather flat
central feature of the data at 3.43 and 3.53 \um\ we have stopped our LR iterative method after 10
iterations, while for the continuum data we have performed a maximum of 100 iterations since no
ringing effect appeared \footnote{Note that stopping the continuum reconstruction at 10 iterations
would not essentially change the result in terms of resulting FWHM (difference less that 5\%).}.

\begin{figure}[htpb]
\includegraphics[angle=0,scale=.50]{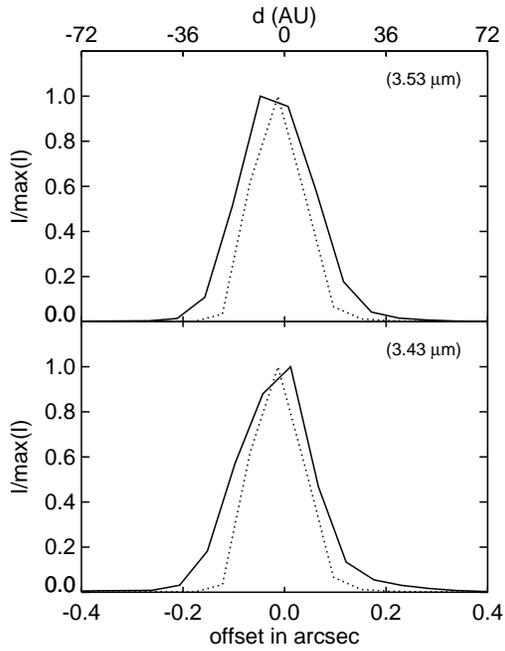}
\caption {Deconvolved normalized emission profiles of the stronger diamond features (continuum substracted,
          solid lines) and of the continuum (dotted lines) obtained using a classical LR method
          (see text for details). The scale on the lower axis shows
the distance in arcsec, that on the upper axis the corresponding value in AU.}
\label{cut_deconvol}
\end{figure}

The results are shown in Fig.\,\ref{cut_deconvol}, from which we derive a full-width at half-maximum
(FWHM) of 0$\farcs$18 (or 32~AU) for the diamond features,
and of 0$\farcs$13 (or 23~AU) for the continuum emission.
These results are consistent with a simple quadratic subtraction assuming
Gaussian light distribution in the pre-processed data.
The signal-to-noise ratio on the spatiale emission profiles is very high (greater than 100)
and the theoretical error on the FWHM from gaussian fits
is less that 1\%. This is clearly an underestimate of the real uncertainties 
of our measurement. Errors on the FWHM values are dominated by systematic 
errors at the level of 10 to 20\%.
The spatial extension derived here for the 3.53 \um\ band is larger by almost a factor of 2 than the
upper limit
inferred by \citet{roche86}. However, the 3.53 \um\ emission feature is spatially unresolved in their
speckle observations performed at the Anglo-Australian 3.9-m telescope (with a diffraction limit
of $\simgreat$0$\farcs$2). The derived upper limit assumes an amount of super-resolution
\citep[see e.g.,][]{Anconelli2004}
that is hardly compatible  with the modest  signal-to-noise ratio  caused by
a seeing value of 3\arcsec\ and no AO correction. On the contrary, the spatial extension we derive here
is approximatively twice  the diffraction-limit of the 8.2-m VLT.
Nevertheless, further accurated AO-assisted 2D imaging with NACO should be performed  to
confirm our size estimates.

\subsubsection{2D intensity maps: the disk is resolved}

Because of the way the slits were positioned and the chop/nod procedure used, we have a 2D coverage of the plane of the sky for a
region of 0.26$\times$4\arcsec, about 47$\times$720~AU. Fig.~\ref{map} shows the map of the
emission of the 3.43 and 3.53 \um\ features and of the continuum.
These maps were obtained by deconvolving independently the data obtained for each of the 7 slit positions,
and hence recomposing the 7 slit vectors into 8 resulting pixels (each slit was shifted by a value that
is half the width of the slit itself during the observations).

Both in the features and in the continuum, the emission is roughly spherical, with FWHM sizes of
about 0.18$\times$0.18\arcsec, 32$\times$32 AU in the features, and 0.13$\times$0.13\arcsec, 23$\times$23 AU in the continuum,
as derived in Sect. 3.2.1 from the analysis of the central slits.
The slight difference between the 3.43 and 3.53 band maps is not significant, given the
uncertainties of the data.

\section{Discussion and conclusions}

Our NACO/VLT observations allow us to resolve for the first time the emission in
the diamond features  (and the adjacent continuum) in a pre-main--sequence star.
The FWHM sizes we derive are typical of emission arising in a circumstellar disk, as we will see in the
following; the roughly spherical shape of the emission maps confirms the hypothesis of a
pole-on disk in HD~97048 put forward by \citet{the86}.


\par\bigskip\noindent
Models of the intensity profile expected from flared circumstellar disks around
Herbig Ae/Be stars, similar to HD~97048, have been recently
computed by \citet{habart2004b}. The models include large grains, in thermal
equilibrium with the stellar radiation field, as well as transiently heated
small graphite grains and PAHs, but not diamonds.
These models show that the continuum emission at $\sim$3 $\mu$m is dominated
by warm large (size of $\sim$0.1 \um) grains at thermal equilibrium; when convolved with a PSF of
FWHM$\sim 0.1 \arcsec$, the intensity profile is only slightly bigger than the PSF, as in our observations.

In all models, the emission of transiently heated particles is significantly broader than that
of the large grains at the same frequency.
This is easily understood when considering the different excitation mechanism, and is discussed
in details in \citet{habart2004b}. In fact, there
is evidence in HD~97048 that the mid-IR PAH emission is extended  on a scale
of 1.2-1.6\arcsec\ or 200-300 AU \citep{boekel03},
comparable  to the outer disk radii seen at millimeter wavelengths in
similar systems \citep{MS97}.
The more extended emission seen by \citet{siebenmorgen2000} must come, instead, from a surrounding nebula.


\par\bigskip\noindent
The diamond emission is  extended more than the continuum emission, but less  than the PAH
emission.
This is roughly consistent with the idea that the diamond carrier of the 3 \um\ features
observed in HD 97048 have sizes intermediate between those of large grains and of PAHs,
 as suggested by detailed analysis of the ISO spectrum  \citep[see e.g.][]{vankerckhoven2002,sheu2002,jones2004}.
Self consistent models which explore the effect of diamond properties
and disk geometry are required to make a useful comparison with observations.

The results presented in this letter add an important piece of information to the
studies of the properties of circumstellar disks and of the nature of the solid particles they contain.
For the first time, we were able
to resolve  the inner disk emission
in the Herbig Ae/Be star HD~97048, to show that the disk is seen pole-on, confirming previous indirect evidence, to resolve the rare emission in the two features at 3.43 and 3.54 $\mu$m
attribute to diamonds and to prove beyond doubts their origin in the inner region
($R \lesssim 15$ AU)  of the disk.

\clearpage
\begin{acknowledgements}
Based on observations collected at the European Southern Observatory, Chile,
ESO N.~70.C-0659.
\end{acknowledgements}

\clearpage



\clearpage

\clearpage

\clearpage

\end{document}